\documentclass[12pt]{article}
\usepackage[dvipdfmx]{graphicx}
\usepackage{float}
\usepackage{amsmath,amssymb}
\begin{document}
\global\arraycolsep=2pt 
\newcommand{\p}{{\bf p}}
\newcommand{\q}{{\bf q}}
\newcommand{\s}{{\bf s}}
\newcommand{\x}{{\bf x}}
\newcommand{\y}{{\bf y}}
%
\thispagestyle{empty} 
\begin{titlepage}    
\begin{flushright}
TNCT-1901
\end{flushright}
\vspace{1cm}
\topskip 3cm
\begin{center}
  \Large{\bf
     Influence of EOS on compact star made of \\
      hidden sector nucleons
                }  \\
\end{center}                                   
\vspace{0.6cm}
\begin{center}
Shinji Maedan            
  \footnote{ E-mail: maedan@tokyo-ct.ac.jp}   
           \\
\vspace{0.8cm}
  \textsl{ Department of Physics, Tokyo National College of Technology,
                    Kunugida-machi,Hachioji-shi, Tokyo 193-0997, Japan}
  \end{center}                                               
\vspace{0.5cm}
\begin{abstract}
\noindent       
\\
We study compact star made of degenerate hidden sector nucleons which
   will be a candidate for cold dark matter.
A hidden sector like QCD is considered, and as the low energy effective
   theory we take (hidden sector) $ SU(2) $ chiral sigma model
   including hidden sector vector meson.
With the mean field approximation, we find that one can treat
   the equation of state (EOS) of our model analytically by introducing a variable
   which depends on the Fermi momentum.
The EOS is specified by the two parameters
   $ C'_{\sigma} $, $ C'_{\omega} $, and we discuss how these parameters affect
   on the mass-radius relation for compact star as well as on the EOS.
The dependence of the maximum stable mass of compact stars on the
   parameter $ C'_{\sigma} $ will also be discussed.
\end{abstract} 
\end{titlepage}
%
%
%
%
%
%
%
\section{Introduction }

Recently many studies have been carried out on compact star
   made of dark matter  \cite{rf:BerHooSil}.
A long time ago, compact stars made of ordinary matter, such as the neutron star,
   have been investigated with the Toleman-Oppenheimer-Volkoff (TOV)
   equations \cite{rf:Tol,rf:OppVol} by which the structure of compact stars involving general
   relativity effects is described.
Since the Universe has an asymmetry between the baryon and anti-baryon
   number density, the neutron star can exist.
In order for the Universe to generate the baryon asymmetry, any required
   mechanism which generates this asymmetry should fulfill the three
   Sakharov conditions (baryon number violation, C and CP violation,
   departure from equilibrium) \cite {rf:Sak}.
Because it seems difficult to generate the baryon asymmetry within the framework
   of the standard model (SM), various mechanisms of baryogenesis have
   been proposed beyond SM.
Here we focus on compact star made of
   dark matter fermions 
   \cite{rf:KhBeBoPuPu,rf:NakMor,rf:NarSchMis,rf:DieLabRaf,rf:GolMohNusRosTep,
            rf:LeuChuLin,rf:HuaXu,rf:XiaJiaZhaYan,rf:MasPniNieKouKok,rf:PanLop,
            rf:WanQiZhaWan,rf:BarBerDel,rf:ChUgEsKo,
            rf:GreZur,rf:WahRahSul}.
For dark matter fermions without any interaction (free dark matter fermions),
   the behavior of the compact star made of degenerate dark matter fermions
   with mass $m_f$ has been made clear.
By taking the unit of length $a$ and the unit of mass $b$,
\begin{equation}
   a \propto \frac{1}{ \sqrt{\gamma} }  \frac{1}{ m^2_f}
                              \frac{1}{ G^{1/2} },  \hskip1cm
   b \propto \frac{1}{ \sqrt{\gamma} }  \frac{1}{ m^2_f}
                              \frac{1}{ G^{3/2} },
  \label{aa}
\end{equation}
   the general character of the solution of the TOV equations is independent of the
   particle properties such as its mass $m_f$ and statistical weight
   $\gamma$ \cite{rf:NakMor,rf:DieLabRaf}.

In this paper, we study compact star made of degenerate dark matter fermions
   interacting with each other.
Among many models predicting dark matter, we take the model similar to
   the one proposed in Ref \cite{rf:HurJunKoLee}, in which 
   the authors consider a hidden sector with a vector like confining gauge
   theory like QCD with $N_{h,c}=3$ colors and $N_{h,f}=2$ flavors.
As the low energy effective theory of this model, we use (hidden sector)
    $ SU(2)_L \times SU(2)_R $ chiral $\sigma$ model \cite{rf:HurJunKoLee,rf:GelLev}
    including hidden sector
    vector meson fields $ {\omega_h}^\mu $ which obtain its mass
    dynamically \cite{rf:Bog}.
This effective theory contains the hidden sector isodoublet
\begin{equation}
    \psi_h = ( p_h, n_h ),
  \label{ab}
\end{equation}
   which we shall call the hidden sector nucleon \cite{rf:HurJunKoLee}.
The lagrangian of this effective theory involves a 'small' term
    $ D \cdot \sigma_h $ which generates nonzero masses for the hidden
    sector pions $\mbox{\boldmath $\pi$}_h $.
Owing to flavor symmetry of the hidden sector, the hidden sector nucleon
   $\psi_h$ and the lightest hidden sector pions
   $\mbox{\boldmath $\pi$}_h $ are both stable, and they will be good candidates for
   cold dark matter \cite{rf:HurJunKoLee}.
Regarding the hidden sector nucleon $\psi_h = ( p_h, n_h )$ as dark matter
   fermions, we shall study characteristic features of the compact star made
   of degenerate hidden sector nucleons, which contains equal numbers of $p_h$ and $n_h$
   (hidden sector isospin symmetric matter).
If the Universe has a symmetry between the hidden sector baryon
   and the hidden sector anti-baryon number density, there is almost
   no possibility to realize these compact stars.
Therefore we assume, without specifying mechanism, the hidden sector
   baryogenesis for which the three Sakharov conditions are demanded
   in the hidden sector (dark sector)
\footnote{
   In the model called asymmetric dark matter (ADM)
   (see  \cite {rf:DavMoh,rf:PetVol,rf:Zur} for review), a mechanism of dark
   baryogenesis is related to that of baryogenesis.
   The models containing composite baryonic dark matter can be found
     in ref \cite {rf:FooVol1,rf:FooVol2,rf:Ber,rf:AlvBehSchWac,
         rf:AnCheMohZha,rf:SpiBehSchWac,rf:Gu1,rf:BucNei,rf:DetMcCPoc,
        rf:Gu2,rf:LonVol,rf:IbeKobNagNak}.
             }
.
The hidden sector isoscalar $ \sigma_h $ will bring attractive
   interaction between the hidden sector nucleons, while
   the hidden sector vector meson $ {\omega_h}^\mu $ will bring repulsive interaction.
To obtain the equation of state (EOS) for the interacting hidden sector nucleons,
   we use the mean field approximation.
Here we assume that when one studies the features of compact star made of
   interacting hidden sector nucleons with the mean field approximation, the
   'small' term $ D \cdot \sigma_h $ in the lagrangian does not play an
   important role.

By the way, before discussing EOS of the hidden sector isospin symmetric
   matter, let us recall EOS of ordinary
   nuclear matter \cite{rf:Kap,rf:PraAin,rf:Gle}.
Consider the system of nucleon, $p$ and $n$,
   involving Yukawa couplings of nucleon to
   scalar meson field $\sigma$ having mass term
   and vector meson field $ \omega^\mu $ having mass term.
The  scalar meson field $\sigma$ will bring attractive interaction between
   nucleons and the vector meson $ \omega^\mu $ will bring 
   repulsive interaction.
With the mean field approximation, the pressure $P$ can be related to the
   energy density $\epsilon$ (EOS) by one independent variable, the Fermi
   momentum $  k_{\rm F} $.
In general it is difficult to express $\epsilon$ or $P$ in explicit analytic form
   of the Fermi momentum $  k_{\rm F} $, because
   in the mean field approximation nucleon has effective mass, and it is hard to
   solve analytically a self-consistent equation for the effective mass.
   
Now we return to the discussion of the hidden sector low energy effective
   theory, and consider the EOS for the interacting hidden sector nucleons with
   the mean field approximation.
We find that in our using model the pressure $P$ can be related to the energy
   density $\epsilon$
   by one variable $\theta$ which depends on the Fermi momentum, and that
   $P$ and $\epsilon$ can be expressed in explicit analytic form of the
   variable $\theta$ introduced.
This fact enables us to study characteristic features of the EOS analytically.
With this EOS which is determined by two parameters
   $  C_{\omega} $ and $ C_{\sigma} $,
   we solve the TOV equations and seek the relation
   between mass $M$ and radius $R$ of the compact star ($M-R$ relation).
It will be interesting to see how the obtained EOS affects on the $M-R$ relation
   for the compact star.
First, the obtained $M-R$ relation for our model including the interaction
   between the hidden sector nucleons will be compared with that for the model of
   a free gas of the hidden sector nucleons whose EOS is well known.
Next, to find the influence of the EOS on the $M-R$ relation for the compact star,
   we vary the value of the dimensionless parameter $ C'_{\sigma} $
   (with fixed dimensionless parameter $ C'_{\omega} $) and examine
   its influence on the $M-R$ relation as well as on the EOS.

This paper is organized as follows.
In Sec.2, the low energy effective theory we use is introduced, and the EOS
   for the interacting hidden sector nucleons
   with the mean field approximation is discussed.
It is emphasized that obtained EOS is given in analytic form by use of
   the variable $\theta$.
In Sec.3, we will discuss characteristic features of the EOS analytically.
In Sec.4, we solve numerically the TOV equations with the EOS, and obtain
   mass $M$ and radius $R$ of compact star made of the interacting hidden
   sector nucleons.
The influence of the EOS on the $M-R$ relation for the compact star
   will be discussed.
Conclusion is given in the last section.
%
%
%
%
%
%
%
%
%
%
\section{The model and EOS}
In this section we introduce the model and discuss the EOS with the
   mean field approximation.
A free gas of the hidden sector nucleon is also treated for later discussion.
\subsection{ The model }
In Ref \cite{rf:HurJunKoLee}, the authors consider a hidden sector with a vector
   like confining gauge theory like QCD.
For the case of $N_{h,c}=3$ colors and $N_{h,f}=2$ flavors, they treat
   (hidden sector) $ SU(2)_L \times SU(2)_R $ chiral sigma model
   as the low energy effective theory, which contains the hidden sector
  isotriplet pions $\mbox{\boldmath $\pi$}_h $, the hidden sector isoscalar $ \sigma_h$,
   and the hidden sector isodoublet $ \psi_h = ( p_h, n_h )$
   called the hidden sector nucleon.
Since the repulsive interaction between the hidden sector nucleons will be important,
   we also include the hidden sector vector meson $ {\omega_h}^\mu $ in the model.
As the low energy effective theory of the hidden sector,
   we shall use the $ SU(2)_L \times SU(2)_R $ chiral sigma model
  including the hidden sector vector meson fields $ {\omega_h}^\mu $
  which obtain its mass dynamically \cite{rf:Bog,rf:SahBasDat,rf:SahOhn},
\begin{eqnarray}
   {\cal L }
  &=& {1 \over 2} \left( \partial_\mu \mbox{\boldmath $\pi$}_h \cdot \,
       \partial^\mu  \mbox{\boldmath $\pi$}_h
       + \partial_\mu \sigma_h \,  \partial^\mu \sigma_h  \right) 
       - {\lambda \over 4} \left(  \mbox{\boldmath $\pi$}_h \cdot \,
              \mbox{\boldmath $\pi$}_h
      + \sigma_h^2 - x^2_0 \right)^2                                              \nonumber \\
  & & \hskip0.1cm  - {1 \over 4} F_{\mu \nu} F^{\mu \nu} + {1 \over 2} g_\omega^2
       \left(  \mbox{\boldmath $\pi$}_h \cdot \, \mbox{\boldmath $\pi$}_h
          + \sigma_h^2 \right)
       {\omega_h}_\mu {\omega_h}^\mu 
       +  \bar\psi_h \left( i \partial \!\!\!/ -g_\omega \gamma_\mu 
                       {\omega_h}^\mu \right) \psi_h                            \nonumber \\
  & &   \hskip0.1cm - g_\sigma \bar{\psi_h} \left( \sigma_h
         + i \gamma_5  \mbox{\boldmath $\tau$ } \cdot \,
               \mbox{\boldmath $\pi$}_h \right) \psi_h - D \cdot \sigma_h,      
    \label{baa}
\end{eqnarray}
where $ \mbox{\boldmath $\tau$} =( \tau_1,  \tau_2,  \tau_3 ) $
   are the Pauli matrices and 
   $ F^{\mu \nu} = \partial^\mu  {\omega_h}^\nu - \partial^\nu  {\omega_h}^\mu $.
The term  $ D \cdot \sigma_h $ originating from the current (hidden sector) quark mass
   breaks the chiral symmetry explicitly and it gives the hidden sector pion field the
   small mass in the vacuum.
Owing to flavor symmetry of the hidden sector, the hidden sector nucleon
   $\psi_h$ and the lightest hidden sector pions $\mbox{\boldmath $\pi$}_h $ are both stable
   and they will be good candidates for cold dark matter \cite{rf:HurJunKoLee}.

In this paper we study the features of compact star made of hidden sector nucleons
   for the system of degenerate hidden sector nucleons.
To obtain field theoretical EOS model, we use the mean field approximation.
We assume that when we study the features of compact star made of hidden sector nucleon
   using the mean field approximation, the small term
   $ D \cdot \sigma_h $ does not play an important role,
   and hereafter we neglect the term $ D \cdot \sigma_h $ in the lagrangian.
When  the term $ D \cdot \sigma_h $ is neglected,
   the vacuum expectation value of the hidden sector $\sigma_h$
   becomes $x_0 (>0) $ \cite{rf:SahOhn},
   and the particle masses of $\psi_h$, $\sigma_h$, and $ {\omega_h}_\mu $ are
\begin{equation}
  m_f
   = g_\sigma x_0,  \hskip1cm  m_\sigma = \sqrt{ 2 \lambda } \, x_0,  \hskip1cm  
  m_\omega = g_\omega x_0,  
  \label{bab}
\end{equation}
respectively.
The equation of motion for $ {\omega_h}_\mu $ in the mean field approximation is
\begin{equation}
    \langle {\omega_h}_0 \rangle =
     \frac{ n_{\rm B} }{ g_\omega x^2 },  \hskip2cm
     \langle {\omega_h}_i \rangle = 0,
  \label{bac}
\end{equation}
where
\begin{equation}
   x = \sqrt{ \langle  \mbox{\boldmath $\pi$}_h \cdot \, \mbox{\boldmath $\pi$}_h
          + \sigma_h^2  \rangle  }.
  \label{bad}
\end{equation}
$ n_{\rm B} $ is the hidden sector baryon density,
\begin{equation}
   n_{\rm B} = \langle {\bar \psi}_h \gamma_0 \psi_h \rangle
          = \frac{ \gamma }{ ( 2 \pi )^3 } \int_0^{ k_{\rm F} } d^3 k,
  \label{bae}
\end{equation}
where $  k_{\rm F} $ is the Fermi momentum and
   $\gamma$ being the statistical factor
   ($ \gamma = 4 $ for isospin doublet).
The equation of morion for $ \sigma_h $ in
   the mean field approximation is \cite {rf:SahBasDat}
\begin{equation}
   y (1-y^2) + \frac{\gamma^2}{18 \pi^4 \, m^2_f \, y^3} \, C_{\sigma} \, C_{\omega}
                \, k^6_{\rm F}
             -\frac{\gamma}{\pi^2} \, C_{\sigma} \, y \, \int^{k_{\rm F}}_{0} d k
             \frac{k^2}{ \sqrt{k^2+m^{*2}_f} } = 0,
  \label{baf}
\end{equation}
where
\begin{equation}
  y \equiv {x / x_0},  \hskip1.5cm
  C_\sigma \equiv g^2_\sigma /m^2_\sigma,   \hskip1.5cm
  C_\omega \equiv g^2_\omega /m^2_\omega,
  \label{bag}
\end{equation}
and $ m^{*}_f \equiv y \, m_f $ is the effective mass of the hidden sector nucleon.
The total energy density $\epsilon$ and
   the pressure $P$ are \cite {rf:SahBasDat}
\begin{eqnarray}
   \epsilon
   &=& \frac{ m^2_f \, (1-y^2)^2 }{8 C_{\sigma} }
          + \frac{\gamma^2}{72 \pi^4 y^2} \, C_{\omega} \, k^6_{\rm F}
          + \frac{\gamma}{ 2 \pi^2} \int^{k_{\rm F}}_{0} d k
              k^2 \sqrt{k^2+ m^{*2}_f },
  \label{bah}
\end{eqnarray}
\begin{eqnarray}
   P
   &=& - \frac{ m^2_f \, (1-y^2)^2 }{8 C_{\sigma} }
          + \frac{\gamma^2}{72 \pi^4 y^2} \, C_{\omega} \, k^6_{\rm F}
          + \frac{\gamma}{ 6 \pi^2} \int^{k_{\rm F}}_{0} d k
             \frac{ k^4 }{ \sqrt{k^2+ m^{*2}_f } }.
  \label{bai}
\end{eqnarray}
Then the EOS is determined when $m_f$, $C_{\omega}$, and $C_{\sigma}$
   are specified.
If the ordinary nuclear matter is considered, the EOS of the nucleon matter with
   the mean field approximation is determined by two parameters $C_{\omega}$
   and $C_{\sigma}$ because the value of the vacuum nucleon mass is
   known, $939 \, {\rm MeV}$ \cite {rf:SahBasDat}.
On the other hand, in our model Eq.(\ref{baa}) the value of the vacuum mass of the hidden
   sector nucleon $m_f$ is unknown.
We now suppose that $m_f$ takes a certain value, and regard $m_f$ as a given constant
   although we do not specify its value here.
According to such assumption the EOS is determined when we specify
   $C_{\omega}$ and $C_{\sigma}$.
It would be useful to define the following dimensionless parameters,
\begin{eqnarray}
   k'_{\rm F} & \equiv & {1 \over m_f} \, k_{\rm F},   \hskip2.3cm
   n'_{\rm B}  \equiv  \frac{ n_{\rm B} }{ m^3_f }
           =  \frac{\gamma}{6 \pi^2} k'^3_{\rm F},                         \nonumber  \\
  C'_{\sigma} & \equiv &   m^2_f C_{\sigma} = \frac{g_{\sigma}^4}{2 \lambda},
                              \hskip1.5cm
  C'_{\omega}  \equiv    m^2_f C_{\omega} = g_{\sigma}^2,             \nonumber  \\
  \epsilon'  & \equiv & \frac{\epsilon}{ m^4_f},   \hskip2.9cm
  P'   \equiv  \frac{P}{ m^4_f}.          \nonumber  \\
  \label{baj}
\end{eqnarray}
In terms of these dimensionless parameters, one has
\begin{eqnarray}
  0 &=& (1-y^2) + \frac{\gamma^2}{18 \pi^4 y^4} \, C'_{\sigma} \, C'_{\omega}
                \, k'^6_{\rm F}
             -\frac{\gamma}{\pi^2} \, C'_{\sigma} \int^{k'_{\rm F}}_{0} d k'
             \frac{k'^2}{ \sqrt{k'^2+y^2} },
  \label{bak}
\end{eqnarray}
\begin{eqnarray}
   \epsilon'
   &=& \frac{ (1-y^2)^2 }{8 C'_{\sigma} }
          + \frac{\gamma^2}{72 \pi^4 y^2} \, C'_{\omega} \, k'^6_{\rm F}
          + \frac{\gamma}{ 2 \pi^2} \int^{k'_{\rm F}}_{0} d k'
              k'^2 \sqrt{k'^2+y^2},
  \label{bal}
\end{eqnarray}
\begin{eqnarray}
   P'
   &=& - \frac{ (1-y^2)^2 }{8 C'_{\sigma} }
          + \frac{\gamma^2}{72 \pi^4 y^2} \, C'_{\omega} \, k'^6_{\rm F}
          + \frac{\gamma}{ 6 \pi^2} \int^{k'_{\rm F}}_{0} d k'
             \frac{ k'^4 }{ \sqrt{k'^2+y^2} }.
  \label{bam}
\end{eqnarray}
The energy per hidden sector nucleon minus hidden sector nucleon mass is
   $ ( \epsilon / n_{\rm B} - m_f ) $ and
\begin{equation}
   { 1 \over m_f } \left( \frac{ \epsilon }{ n_{\rm B} } - m_f  \right)
      = \frac{ \epsilon' }{ n'_{\rm B} } - 1.
  \label{ban}
\end{equation}
\subsection{EOS in terms of a parameter $ \theta $}
Usually one often takes the dimensionless Fermi momentum $ k'_{\rm F} $
   as the one independent variable in the three equations
   (\ref{bak}), (\ref{bal}), and (\ref{bam}).
By solving self-consistently the equation Eq.(\ref{bak}) for $y$, both $ \epsilon' $
   and $ P' $ are expressed by the use of the one independent variable $ k'_{\rm F} $,
   thus one obtains the EOS within the mean field approximation.
   
In this paper we shall use another technique.
The equation of motion for $\sigma_h$ can be expressed as
\begin{eqnarray}
  0 &=& {1 \over y^2} - 1 + \frac{\gamma^2}{18 \pi^4 y^6} \, C'_{\sigma} \,
                  C'_{\omega} \, k'^6_{\rm F}
            -\frac{\gamma}{\pi^2} C'_{\sigma} \int^{ k'_{\rm F}/y }_{0} d \phi
             \frac{ \phi^2}{ \sqrt{\phi^2+1} }.
  \label{bba}
\end{eqnarray}
We here introduce a variable $ \theta $ defined by
\begin{equation}
   \theta \equiv \frac{ k'_{\rm F} }{y} \ge 0,
  \label{bbb}
\end{equation}
which can also be written as $ \theta= k_{\rm F} / m^{*}_f $
   \cite{rf:WahRahSul}.
The equation of motion then has the form,
\begin{equation}
   {1 \over y^2}
     = 1 - \frac{\gamma^2}{18 \pi^4} \, C'_{\sigma} \,
                  C'_{\omega} \, \theta^6
            + \frac{\gamma}{\pi^2} C'_{\sigma}  \int^{ \theta }_{0} d \phi
             \frac{ \phi^2}{ \sqrt{\phi^2+1} }.
       \label{bbc}
\end{equation}
If we define a function $ f(\theta) $ by the right-handed side of Eq.(\ref{bbc}),
\begin{eqnarray}
      f(\theta) &\equiv&
         1 - \frac{\gamma^2}{18 \pi^4} \, C'_{\sigma} \,
                  C'_{\omega} \, \theta^6
            + \frac{\gamma}{\pi^2} C'_{\sigma}  \int^{ \theta }_{0} d \phi
             \frac{ \phi^2}{ \sqrt{\phi^2+1} }             \nonumber \\
       &=&  1 + C'_{\sigma} \times  \gamma  \left[ - \frac{\gamma }{18 \pi^4}
                 \, C'_{\omega} \, \theta^6
            + \frac{ 1 }{ 2 \,\pi^2}  
                 \{ \theta \sqrt{ \theta^2+1 }
                       - {\rm arcsinh} \, \theta \}   \right],
    \label{bbd}
\end{eqnarray}
it has the following features.
The equation $ f( \theta ) =0 $ has a solution $ \theta_f (>0)$ whose value is determined
   by $  C'_{\omega} $ and $ C'_{\sigma} $, and $ f( \theta ) $ has the following property,
\begin{equation}
    f( \theta )  \left\{  \begin{array}{ll}
                                 > 0   &  ( 0 \le \theta < \theta_f  )   \\
                                 < 0    &  (  \theta_f < \theta ) 
                              \end{array}
                    \right.  
    \label{bbe}
\end{equation}
For a given value of  $  k'_{\rm F} $, the unknown quantity $\theta$ is
   obtained by solving the equation of motion,
\begin{equation}
 {1 \over  k'^2_{\rm F} } \, \theta^2 =  f( \theta ),
  \label{bbf}
\end{equation}
   where $\theta$ should satisfy $ 0 \le \theta < \theta_f $.
Using this solution, one can calculate $y$ by $  y= 1 / \sqrt{ f( \theta ) }$.
$ \epsilon' $ and $ P' $ are expressed in terms of $ \theta $,
\begin{eqnarray}
    \epsilon'
    &=&   \frac{1}{ f(\theta)^2 } \biggl[  \frac{\gamma}{2 \pi^2} 
               \int^{ \theta }_{0} d \phi \, \phi^2  \sqrt{\phi^2+1} 
                                   \biggr.                      \nonumber  \\
   & &  \hskip1.8cm   \biggl.  + \frac{ ( f(\theta) - 1 )^2 }{8 C'_{\sigma} }
         + \frac{\gamma^2}{72 \pi^4} \, C'_{\omega} \, \theta^6
            \biggr]                                            \nonumber  \\
   &=& \frac{1}{ f(\theta)^2 } \left[
             \frac{\gamma}{16 \pi^2 } \biggl\{ \theta \, (2 \theta^2 + 1) \sqrt{ \theta^2+1 }
                  - \, {\rm arcsinh} \, \theta \right\}         \biggr.                      \nonumber  \\
   & &  \hskip1.8cm   \biggl.  + \frac{ ( f(\theta) - 1 )^2 }{8 C'_{\sigma} }
         + \frac{\gamma^2}{72 \pi^4} \, C'_{\omega} \, \theta^6
            \biggr],
  \label{bbg}
\end{eqnarray}
\begin{eqnarray}
    P'
    &=&   \frac{1}{ f(\theta)^2 } \biggl[  \frac{\gamma}{6 \pi^2} 
               \int^{ \theta }_{0} d \phi \,
                  \frac{ \phi^4}{ \sqrt{\phi^2+1} } 
                                   \biggr.                      \nonumber  \\
   & &  \hskip1.8cm   \biggl.  - \frac{ ( f(\theta) - 1 )^2 }{8 C'_{\sigma} }
         + \frac{\gamma^2}{72 \pi^4} \, C'_{\omega} \, \theta^6
            \biggr]                                             \nonumber  \\
   &=& \frac{1}{ f(\theta)^2 } \left[
             \frac{\gamma}{48 \pi^2 } \biggl\{ (2 \theta^3 - 3 \theta ) \sqrt{ \theta^2+1 }
                  +3 \, {\rm arcsinh} \, \theta \right\}         \biggr.                      \nonumber  \\
   & &  \hskip1.8cm   \biggl.  - \frac{ ( f(\theta) - 1 )^2 }{8 C'_{\sigma} }
         + \frac{\gamma^2}{72 \pi^4} \, C'_{\omega} \, \theta^6
            \biggr],
  \label{bbh}
\end{eqnarray}
   thus the dimensionless total energy density $ \epsilon' $ and  dimensionless
   pressure $ P' $ can be expressed by only one variable
   $ \theta \, ( 0 \le \theta < \theta_f ) $, respectively.
If the variable $\theta$ can be eliminated, one can obtain the (dimensionless) equation
   of state (EOS), which, apart from the statistical factor $\gamma$, is determined when the
   two dimensionless parameters $ C'_{\sigma} $ and $ C'_{\omega} $ are given.
The dimensionless hidden sector baryon number  $ n'_{\rm B} $ is also
   expressed in terms of $\theta$,
\begin{equation}
  n'_{\rm B} = \frac{\gamma}{6 \pi^2} k'^3_{\rm F}
                  = \frac{\gamma}{6 \pi^2} \frac{ \theta^3 }{ f( \theta )^{3/2} }.
  \label{bbi}
\end{equation}
\subsection{ A free gas of hidden sector nucleon }
In order to make the features of our EOS clear, it will be helpful to review
   the EOS of a free gas of fermion \cite{rf:NarSchMis}.
We consider a free gas of hidden sector nucleon ($\gamma=4$) with mass $m_f$
   which is the same magnitude as the vacuum mass of the hidden sector
   nucleon Eq.(\ref{bab}).
The dimensionless total energy density $ \epsilon' \equiv \epsilon / m_f^4 $ and
   dimensionless pressure $ P' \equiv P / m_f^4 $ are given by
\begin{eqnarray}
    \epsilon'
   &=&  \frac{\gamma}{ 2 \pi^2} \int^{k'_{\rm F}}_{0} d k'
              k'^2 \sqrt{k'^2+1}                       \nonumber \\
   &=&   \frac{\gamma}{16 \pi^2 } \biggl\{ k'_{\rm F} \, (2 \, k'^2_{\rm F} + 1)
              \sqrt{ k'^2_{\rm F} +1 }
                  - \, {\rm arcsinh} \, k'_{\rm F}  \biggr\},  
    \label{bca}
\end{eqnarray}
\begin{eqnarray}
    P'
   &=&  \frac{\gamma}{ 6 \pi^2} \int^{k'_{\rm F}}_{0} d k'
             \frac{ k'^4 }{ \sqrt{k'^2+1} }                \nonumber \\
   &=&   \frac{\gamma}{48 \pi^2 } \biggl\{(2 \, k'^3_{\rm F} - 3 \,  k'_{\rm F} )
              \sqrt{ k'^2_{\rm F} +1 }
                  +3 \, {\rm arcsinh} \, k'_{\rm F}  \biggr\}.
    \label{bcb}
\end{eqnarray}
In the nonrelativistic case $ k'_{\rm F} \ll 1$, one has
   $  \epsilon' \approx ({\gamma}/{6 \pi^2 }) k'^3_{\rm F} $,
   $  P' \approx ({\gamma}/{30 \pi^2 }) k'^5_{\rm F} $,
   and finds the well known relation,
\begin{equation}
   P' \approx {1 \over 5} \left(  \frac{6 \pi^2}{\gamma} 
       \right)^{2 \over 3}  {\epsilon'}^{5 \over 3}
        \propto {\epsilon'}^{ 5/3}.
        \label{bcd}
\end{equation}
At high densities $ k'_{\rm F} \gg 1$, one has
   $ \epsilon' \approx  ({\gamma}/{8 \pi^2}) k'^4_{\rm F} $,
   $ P' \approx ({\gamma}/{24 \pi^2 }) k'^4_{\rm F} $,
   and finds the well known relation,
\begin{equation}
    { P' \over  \epsilon' } \approx {1 \over 3}.
  \label{bcf}
\end{equation}
We finally consider the case when $ k'_{\rm F} = 1$
   or $ k_{\rm F} = m_f $ ($\gamma=4$),
\begin{equation}
    \epsilon'
          =  \frac{\gamma}{ 2 \pi^2} \int^{1}_{0} d k'
              k'^2 \sqrt{k'^2+1} 
         \approx 0.08514,
    \label{bcg}
\end{equation}
\begin{equation}
    P'
          = \frac{\gamma}{ 6 \pi^2} \int^{1}_{0} d k'
             \frac{ k'^4 }{ \sqrt{k'^2+1} }
          \approx 0.01038.
    \label{bch}
\end{equation}
Then at $ k'_{\rm F} = 1$ one has
\begin{equation}
   \frac{P'}{\epsilon'} \approx 0.1219.
  \label{bci}
\end{equation}
%
%
%
%
%
%
\section{ Characteristic features of the EOS }
In the preceding section, we have seen that the relation between the energy density
   $\epsilon'$ and the pressure $P'$ can be expressed by the use of only
   one parameter $\theta$.
It should be pointed out that $\epsilon'$ and $P'$ are represented by explicit
   functions of $\theta$, respectively,
   so that we can study characteristic features of the EOS analytically.
For the definiteness, we consider the case of $ C'_{\sigma}  >  C'_{\omega} $
  in this section and the next section.
Eqs.(\ref{bbg}) and (\ref{bbh}) are written more symmetric form,
\begin{equation}
   \epsilon'
      =   \frac{1}{ f(\theta)^2 } \biggl[  \frac{\gamma}{2 \pi^2} 
               \int^{ \theta }_{0} d \phi \, \phi^2  \sqrt{\phi^2+1} 
         +  \frac{\gamma^2}{72 \pi^4} \, \left\{ C'_{\omega}
                          +  g(\theta)^2  C'_{\sigma} \right\}  \theta^6
                                                    \biggr],
      \label{cxa}
\end{equation}
\begin{equation}
    P'
     =   \frac{1}{ f(\theta)^2 } \biggl[   \frac{\gamma}{6 \pi^2} 
               \int^{ \theta }_{0} d \phi \,
                  \frac{ \phi^4}{ \sqrt{\phi^2+1} } 
          +  \frac{\gamma^2}{72 \pi^4} \, \left\{ C'_{\omega}
                          -  g(\theta)^2  C'_{\sigma} \right\}  \theta^6
                           \biggr],
     \label{cxb}
\end{equation}
   where we have defined $g( \theta )$,
\begin{eqnarray}
   g( \theta ) 
   & \equiv &   \frac{3 \pi^2}{\gamma  C'_{\sigma} } 
                            \frac{ ( f(\theta) - 1 ) }{ \theta^3 }                          \nonumber  \\
   & = & \frac{3}{ \theta^3 }   \int^{ \theta }_{0} d \phi  \frac{ \phi^2}{ \sqrt{\phi^2+1} }  
             -  \frac{\gamma}{ 6 \pi^2}  C'_{\omega} \, \theta^3,
  \label{cxc}
\end{eqnarray}
  which does not depend on $ C'_{\sigma} $.
For any $ C'_{\omega} >0 $ the $g(\theta)$ has the following properties,
   $ g(\theta) \rightarrow 1 \, (\theta \rightarrow 0)$, and
   $ d \, g(\theta) / d \, \theta <0 $, hence $ g(\theta) $ is a monotone decreasing function
   of $\theta$ and satisfies $ g(\theta) < 1 $.
   
For convenience we shall call the first term in a bracket of the right-handed side
   of Eq.(\ref{cxa}) 'free term', and the second term 'interaction term'.
Properly speaking, the first term in Eq.(\ref{cxa}) involves  interaction effects through $y$
   (because $\theta = k'_{\rm F} /y$).
In the same way we shall call the  first term in a bracket of the right-handed side
   of Eq.(\ref{cxb}) 'free term', and the second term 'interaction term'.
Let us pay attention to the term
$  \left\{ \, C'_{\omega}  -  g( \theta )^2  C'_{\sigma} \right\} $
in the interaction term of the dimensionless pressure $P'$.
This term can take negative values or positive values according to $ g(\theta)$,
\begin{equation}
     \left\{ \, C'_{\omega}  -  g( \theta )^2  C'_{\sigma} \right\}
         \left\{  \begin{array}{ll}
                      < 0                   &  ( \sqrt{ {C'_{\omega} \over C'_{\sigma}} } < g(\theta) < 1)   \\
                      \ge 0                &  ( 0 < g(\theta) <  \sqrt{ {C'_{\omega} \over C'_{\sigma}} } ) \\
                      = C'_{\omega} &  ( g(\theta)=0 )
                    \end{array}
         \right.
  \label{cxd}
\end{equation}
In the limit $\theta \rightarrow 0,  g(\theta) \rightarrow 1 $ and
   this term has a negative value,
   $C'_{\omega} - C'_{\sigma}<0$.
Because $ g(\theta) $ is a monotone decreasing function of $\theta$,
   as $\theta$ becomes larger, the term
   $ \left\{ \, C'_{\omega}  -  g( \theta )^2  C'_{\sigma} \right\} $
   in the pressure $P'$ changes over from negative to positive.
The \linebreak changeover point of the sign of
   $  \left\{ \, C'_{\omega}  -  g( \theta )^2  C'_{\sigma} \right\} $
   is $ g(\theta) = \sqrt{  C'_{\omega} / C'_{\sigma} } $.
This term  \linebreak $  \left\{ \, C'_{\omega}  -  g( \theta )^2  C'_{\sigma} \right\} $
   in the interaction term of $P'$ will play an important role in making the EOS softer or stiffer,
   as will be discussed later.
Note that when $ g(\theta)=0$, we have $f(\theta)=1$ and both $\epsilon'$ and $P'$ do not
   depend on $C'_{\sigma}$ at the point $\theta$ which satisfies $ g(\theta)=0$.
This point will be discussed in detail in section 4.
\subsection{ Nonrelativistic case, $ k'_{\rm F} \ll 1$}
For a given value of  $  k'_{\rm F} $, the quantity $\theta$ is
   obtained by solving $  \theta^2 / k'^2_{\rm F} = f(\theta) $.
In the nonrelativistic region, $  k'_{\rm F} \ll 1 $, the $\theta$ becomes $ \theta \ll 1 $
   and we have
\begin{equation}
   f(\theta) 
      =  1+ \frac{\gamma}{3 \pi^2} \, C'_{\sigma} \, \theta^3
                - \frac{\gamma}{10 \pi^2} \, C'_{\sigma} \, \theta^5
                - \frac{\gamma^2}{18 \pi^4} \, C'_{\sigma} \, C'_{\omega} \, \theta^6
               + O(\theta^7)>1, 
     \label{caa}
\end{equation}
\begin{equation}
  g( \theta )   
        =   \left\{ 1 - {3 \over 10} \theta^2 +  {9 \over 56} \theta^4
                                   + O( \theta^6 ) \right\}
                                   -   \frac{\gamma}{ 6 \pi^2}  C'_{\omega} \, \theta^3 <1,
  \label{cab}
\end{equation}
   hence $y$ is less than $1$, $ y=1/ {\sqrt{ f( \theta ) } }<1$.
$\epsilon'$ and $P'$ in the nonrelativistic region become
\begin{eqnarray}
  \epsilon'
   &=&   \frac{1}{ f(\theta)^2 } \frac{\gamma}{16 \pi^2 } \left\{ {8 \over 3} \, \theta^3
            + O(\theta^5) \right\} + \frac{\gamma^2}{72 \pi^4} \left( C'_{\omega}
            + C'_{\sigma} \right) \theta^6 + O(\theta^8),                        \nonumber   \\   
     \label{cac}
\end{eqnarray}
\begin{eqnarray}
   P'
    &=&   \frac{\gamma}{48 \pi^2 } \left\{ {8 \over 5} \, \theta^5
            + O(\theta^7) \right\}
            +  \frac{\gamma^2}{72 \pi^4} \left( C'_{\omega} -  C'_{\sigma} \right) \theta^6
            + O(\theta^8).                        \nonumber   \\
     \label{cad}
\end{eqnarray}
In the leading order of $ \theta $, the interaction terms can be neglected
   and we find the relation
\begin{equation}
   P' \rightarrow {1 \over 5} \left(  \frac{6 \pi^2}{\gamma}
       \right)^{2 \over 3} {\epsilon'}^{5 \over 3} 
       \hskip1cm  (\theta \rightarrow 0).
        \label{cae}
\end{equation}
Therefore in the nonrelativistic limit, $  k'_{\rm F} \rightarrow 0 \, ( \theta \rightarrow 0 ) $,
   the EOS is the same as that of the free theory.

Next, we shall consider the contribution of the interaction term
   $ ( C'_{\omega} -  C'_{\sigma} ) \theta^6 $ in Eq.(\ref{cad}) which
   is negative because of $ C'_{\omega} < C'_{\sigma} $.
Since the contribution of the interaction term to the free term in the pressure
   is much larger than that in the energy density, EOS becomes softer compared with
   the free theory.
When $ \vert  C'_{\omega} -  C'_{\sigma} \vert $ becomes larger, the EOS becomes softer.
In the next leading order of $ \theta $ $( 0 < \theta \ll 1 )$,
\begin{equation}
  \frac{ d ( \log_{10} P' ) }{ d ( \log_{10} \epsilon' ) }
  = \frac{ \epsilon' }{ P' }  \frac{ ( d P' / d \theta ) }{  ( d \epsilon' / d \theta ) }
  = {5 \over 3} \left[ 1+ \frac{ \gamma }{ 12 \pi^2 } 
        \left( C'_{\omega} -  C'_{\sigma} \right) \theta +O( \theta^2 ) \right].
  \label{caf}
\end{equation}
We can say that at the neighborhood of a small parameter value $\theta$,
   $P'$ and $\epsilon'$ will be related by
\begin{equation}
    P' \propto \epsilon'^{ 
       {5 \over 3} \left[ 1+ \frac{ \gamma }{ 12 \pi^2 } 
        \left( C'_{\omega} -  C'_{\sigma} \right) \theta \right] },
  \label{cag}
\end{equation}
where 
$ (5 / 3) \left[ 1+ ( \gamma / 12 \pi^2 ) 
        \left( C'_{\omega} -  C'_{\sigma} \right) \theta \right]  <  (5 / 3) $
for $  C'_{\omega} < C'_{\sigma}  $.
\subsection{At high densities, $ k'_{\rm F} \gg 1$}
At extremely high density, $ k'_{\rm F} \rightarrow \infty $, the chiral sigma model
   will not work since the chiral sigma model is one of the low energy effective theory of QCD.
Although we should use the hidden QCD theory in such region, it will be interesting to study
   the model Eq.(\ref{baa}) at high densities, $ k'_{\rm F} \gg 1$  \cite{rf:Bog}.
Note that in our model Eq.(\ref{baa}) the hidden vector meson mass is generated dynamically.
For a given value of  $  k'_{\rm F} $, the unknown $\theta$ is
   obtained by solving $  \theta^2 / k'^2_{\rm F} = f(\theta) $.
At high densities, $  k'_{\rm F} \gg 1 $, $\theta$ becomes
\begin{equation}
               \theta \approx \theta_f,
  \label{cba}
\end{equation}
   where $\theta_f ( >0)$ is a solution of the equation $ f(\theta) = 0 $ and can be determined
   by $  C'_{\omega} $ and $  C'_{\sigma}  $, not depending on $ k'_{\rm F} $.
The value of $y$ then becomes
\begin{equation}
  y =  { k'_{\rm F} \over \theta} \approx {1 \over \theta_f} \, k'_{\rm F} \gg 1,
  \label{cbb}
\end{equation}
   which means that the effective mass of the hidden sector nucleon
   $ m^{*}_f \rightarrow \infty$ when $ n_{\rm B} \rightarrow \infty$.
In other words chiral symmetry is not restored for asymptotic densities
   in the model Eq.(\ref{baa}) \cite{rf:Bog}.

We study EOS of our model at high densities $ k'_{\rm F} \gg 1$.
One finds
\begin{eqnarray}
  \epsilon'
   & \approx &   k'^4_{\rm F} \times {1 \over \theta_f^4} 
             \frac{\gamma}{2 \pi^2 }  \int^{ \theta_f }_{0} d \phi \,  \phi^2  \sqrt{\phi^2+1}
                                                                                      \nonumber   \\
     & &  \hskip1.9cm   +   k'^4_{\rm F} \times {1 \over \theta_f^4} 
             \left\{  \frac{ 1 }{8 C'_{\sigma} }
               +  \left( \frac{\gamma}{6 \pi^2} \right)^2
                  \left( \frac{ C'_{\omega} }{2} \right) \, \theta_f^6  \right\},
  \label{cbc}
\end{eqnarray}
\begin{eqnarray}
   P'
   & \approx &    k'^4_{\rm F} \times {1 \over \theta_f^4}
             \frac{\gamma}{6 \pi^2 }   \int^{ \theta_f }_{0} d \phi  \frac{ \phi^4}{ \sqrt{\phi^2+1} }
                                                                                            \nonumber   \\
   & & \hskip1.9cm         +  k'^4_{\rm F} \times {1 \over \theta_f^4}
           \left\{ - \frac{ 1 }{8 C'_{\sigma} }
             +  \left( \frac{\gamma}{6 \pi^2} \right)^2
                  \left( \frac{ C'_{\omega} }{2} \right) \, \theta_f^6  \right\}.
  \label{cbd}
\end{eqnarray}
Both the free term and the interaction term contribute the 
   same order $ O(  k'^4_{\rm F} ) $, and consequently $ P'/ \epsilon'$ becomes
\begin{equation}
  { P' \over \epsilon' } \approx
  \frac{
            \biggl[
             \frac{\gamma}{6 \pi^2 }   \int^{ \theta_f }_{0} d \phi
             \frac{ \phi^4}{ \sqrt{\phi^2+1} }  - \frac{ 1 }{8 C'_{\sigma} }
            +  \left( \frac{\gamma}{6 \pi^2} \right)^2
                  \left( \frac{ C'_{\omega} }{2} \right) \, \theta_f^6      \biggr]   }
         {
             \biggl[
             \frac{\gamma}{2 \pi^2 }  \int^{ \theta_f }_{0} d \phi \,
              \phi^2  \sqrt{\phi^2+1}  + \frac{ 1 }{8 C'_{\sigma} }
            +  \left( \frac{\gamma}{6 \pi^2} \right)^2
                  \left( \frac{ C'_{\omega} }{2} \right) \, \theta_f^6        \biggr]     }
     < 1,
      \label{cbe}
\end{equation}
   for $ k'_{\rm F} \gg 1$.
The above right-handed side is a constant whose value is determined by
   $  C'_{\omega} $ and $  C'_{\sigma}  $, because $\theta_f $ can be expressed
   by  $  C'_{\omega} $ and $  C'_{\sigma}  $.
If the interaction terms are dominant in Eq.(\ref{cbe}), we have
\footnote{
   The interaction term
     $  \left\{ - \frac{ 1 }{8 C'_{\sigma} }
         +  \left( \frac{\gamma}{6 \pi^2} \right)^2
          \left( \frac{ C'_{\omega} }{2} \right) \, \theta_f^6  \right\}  $
   in $P'$ can be shown to be positive with the help of the definition
    $ f( \theta_f )=0 $.
                                   }
\begin{equation}
  { P' \over \epsilon' } \approx
  \frac{
            \biggl[
            - \frac{ 1 }{8 C'_{\sigma} }
            +  \left( \frac{\gamma}{6 \pi^2} \right)^2
                  \left( \frac{ C'_{\omega} }{2} \right) \, \theta_f^6      \biggr]   }
         {
             \biggl[
            \frac{ 1 }{8 C'_{\sigma} }
            +  \left( \frac{\gamma}{6 \pi^2} \right)^2
                  \left( \frac{ C'_{\omega} }{2} \right) \, \theta_f^6        \biggr]     }
     < 1.
      \label{cbg}
\end{equation}
If the interaction terms are negligible, we have
\begin{equation}
  { P' \over \epsilon' } \approx
          \frac{  \left[ \frac{\gamma}{6 \pi^2 } 
                             \int^{ \theta_f }_{0} d \phi \frac{ \phi^4}{ \sqrt {\phi^2+1} } \right]   }
                 {  \left[
                  \frac{\gamma}{2 \pi^2 }  
                             \int^{ \theta_f }_{0} d \phi \, \phi^2  \sqrt{\phi^2+1} \right]    }
    < {1 \over 3},
       \label{cbh}
\end{equation}
   because of simply verified relation
\begin{equation}
    \int^{ \theta_f }_{0} d \phi \,  \phi^2  \sqrt{\phi^2+1}
    >    \int^{ \theta_f }_{0} d \phi
             \frac{ \phi^4}{ \sqrt{\phi^2+1} }.
  \label{cbf}
\end{equation}
Note that in a free gas model we know
   $ P' / \epsilon'  \approx 1/3 $
   at high densities $ k'_{\rm F} \gg 1$.
%
%
%
%
%
%
%
\section{ TOV equation and compact star}
In this section we solve the TOV equations with the EOS numerically. 
The statistical weight is $\gamma=4$ for isospin doublet and, as noted in section 3,
   we set  $  C'_{\omega} < C'_{\sigma} $  in the preceding section and this section.
In order to do the numerical calculations, we should set the values of the
  dimensionless parameters $  C'_{\omega} $ and $  C'_{\sigma} $.
We choose in this section the parameters  $  C'_{\omega} $ and $  C'_{\sigma} $
   so as to be $ y=1 $ when $ k'_{\rm F} $ takes a value
   $1$, i.e., $ \theta = 1 $ when $ k'_{\rm F} =1 $.
This leads to
\begin{equation}
   1  =  f( \theta =1 )                        
       =  1 + \gamma  C'_{\sigma} \, \left[ 
          - \frac{\gamma}{18 \pi^4} \, C'_{\omega}
            +  {1 \over 2 \pi^2 } \{ \sqrt{ 2 } - {\rm arcsinh} ( 1 ) \}
                 \right],                      
     \label{dxa}
\end{equation}
and then $ C'_{\omega} $ becomes
\begin{equation}
        C'_{\omega} = \frac{9 \pi^2}{\gamma} 
                                \left\{ \sqrt{ 2 } - {\rm arcsinh} ( 1 ) \right\}
                             \approx 11.8326  \hskip1cm ( \gamma=4 ).
    \label{dxb}
\end{equation}
With this value of $ C'_{\omega} $, we can show that if $ 0 <  k'_{\rm F} < 1$, then $ y < 1 $,
   and if $ 1 <  k'_{\rm F} $, then $ 1 < y  $ (The proof will be found in Appendix A).
The requirement $ f(\theta=1) = 1 $, however, does not restrict the value of $ C'_{\sigma} $,
   therefore we will consider in this paper the parameter region of $  C'_{\sigma} $ in which
   the system becomes homogeneous and a bound state of hidden sector nucleon
   does not appear.
If one restrict as  $  C'_{\sigma} \lesssim 30 $, the conditions
   $ \partial P / \partial n_{\rm B} >0 $ and $ \epsilon / n_{\rm B} -m_f > 0$ are satisfied
   when the above value Eq.(\ref{dxb}) of  $C'_{\omega}$ is taken.
With this value of $  C'_{\omega} $ and $  C'_{\sigma} \lesssim 30 $, one can ascertain that
   the quantity $ \theta^2 /f(\theta) $ becomes
  a monotone increasing function of $\theta$.
This ascertainment leads to the statement that $\theta$ is a monotone increasing function
   of $ k'_{\rm F} $ since the unknown quantity $\theta$ is obtained by solving the equation
   of motion $ k'^2_{\rm F} = \theta^2 / f(\theta) $.

With our choice of the value of $  C'_{\omega} $ which leads to
   $ f(\theta =1) =1 $, the EOS at the point $ \theta=1$ does not depend on
   $  C'_{\sigma} $.
This can be seen as follows.
At the point $ \theta = 1 $, one has 
   $  g( \theta ) = 3 \pi^2 ( f(\theta) - 1 ) / \gamma C'_{\sigma} \theta^3 =0$,
   so that $ C'_{\omega} \pm g(\theta)^2 \, C'_{\sigma} = C'_{\omega} $
   and both $\epsilon'$ and $P'$ do not depend on $  C'_{\sigma} $ at the point $ \theta=1$.
Thus we find that the EOS does not depend on $  C'_{\sigma} $ at the point $ \theta = 1 $.
Further we shall examine whether the EOS of our model at $\theta=1$ is stiffer or
   softer than the EOS of a free gas model, because we can discuss well
   such a question analytically when $\theta=1$.
In our model we have at $\theta=1$
\begin{eqnarray}
   \epsilon'
    &=&   \frac{\gamma}{2 \pi^2} 
               \int^{ 1 }_{0} d \phi \, \phi^2  \sqrt{\phi^2+1} 
               + \frac{\gamma^2}{72 \pi^4} \, C'_{\omega}
    \approx  0.1121358,                   \nonumber   \\
    P'
   &=&   \frac{\gamma}{6 \pi^2} 
               \int^{ 1 }_{0} d \phi \,
                  \frac{ \phi^4}{ \sqrt{\phi^2+1} } 
             + \frac{\gamma^2}{72 \pi^4} \, C'_{\omega} 
   \approx  0.0373786,
       \label{dxc}
\end{eqnarray}
and
\begin{equation}
   \frac{ P'}{\epsilon'} \approx 0.33333,  \hskip1.2cm
    (  \log_{10} \epsilon',   \log_{10} P' ) \approx ( -0.950, -1.427 ).
  \label{dxd}
\end{equation}
In a free gas model the value $0.1121358$ of the energy density is obtained when
   $ k'_{\rm F} =1.08477 $ by numerical calculation, 
   $  \epsilon'( k'_{\rm F}=1.08477) = 0.1121358$.
We introduce $\Delta \epsilon'$ and  $\Delta P'$ in a free gas model by the following,
\begin{eqnarray}
    \epsilon'( k'_{\rm F}=1.08477)
       & = & \frac{\gamma}{ 2 \pi^2} \int^{1}_{0} d k'
              k'^2 \sqrt{k'^2+1} + \Delta \epsilon'
          =0.1121358,                             \nonumber \\
     P'( k'_{\rm F}=1.08477)
       &  = & \frac{\gamma}{ 6 \pi^2} \int^{1}_{0} d k'
             \frac{ k'^4 }{ \sqrt{k'^2+1} } + \Delta P'.          
    \label{dxe}
\end{eqnarray}
One has
\begin{equation}
  \frac{ \Delta P'}{ \Delta \epsilon'} \approx
    \frac{d P'}{ d\epsilon'}  \biggr |_{ k'_{\rm F} =1} =
      {1 \over 3} \frac{ k'^2_{\rm F} }{ (k'^2_{\rm F}+1) }
                      \biggr |_{ k'_{\rm F} =1} =
    {1 \over 6},
  \label{dxf}
\end{equation}
hence
\begin{equation}
    \Delta P' \approx {1 \over 6} \Delta \epsilon'.
  \label{dxg}
\end{equation}
Now let us compare $\epsilon'$ and $P'$ in our model Eq.(\ref{dxc}) 
   with those in a free gas model Eq.(\ref{dxe}).
Note that the first term of the r.h.s of $\epsilon' ( P' )$ in Eq.(\ref{dxc}) is the same as
   that of $\epsilon' ( P' )$ in Eq.(\ref{dxe}).
In our model the same quantity $ (\gamma^2 / {72 \pi^4} ) \, C'_{\omega} $ is
   added to both the first term of the r.h.s of $\epsilon'$ and that of $P'$.
On the other hand, in a free gas model  $\Delta P'$ added to the first term of the r.h.s of $P'$
   is much smaller than $ \Delta \epsilon'$ added to the first term of the r.h.s
   of $\epsilon'$.
We thus see that the pressure at $\theta=1$ of our model is larger than that of a free
   gas model for the same energy density value, and conclude that  the EOS of our model at
   $\theta=1$ is stiffer than that of a free gas model.
This will be ensured later by numerical calculations of EOS.
\subsection{ Numerical calculations of EOS with
               $  C'_{\sigma} = (4/3) C'_{\omega},  (5/3) C'_{\omega}, (6/3) C'_{\omega} $.  }
%
%
%
%
%
%
\begin{figure}
     \centering
     \includegraphics[height=8.7cm]{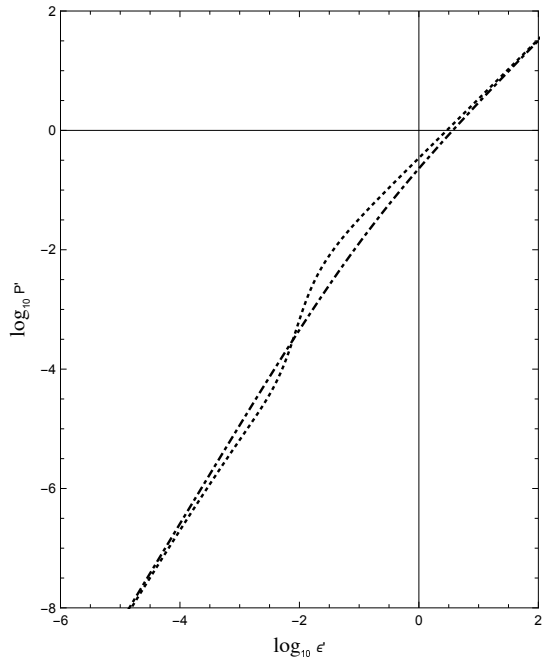}
    \caption{The graphs of EOS (dimensionless energy density $\epsilon'$ and
                  dimensionless pressure $P'$) with $ C'_{\sigma} = (6/3) C'_{\omega}$ (dotted)
                  in addition to a free gas case (dash-dotted).
                  We take $C'_{\omega} = (9 \pi^2 / 4 ) \left\{ \sqrt{ 2 } - {\rm arcsinh} ( 1 ) \right\}$.}
       \label{fig:1}
\end{figure}
We have set
   $  C'_{\omega} = (9 \pi^2 / 4 ) \left\{ \sqrt{ 2 } - {\rm arcsinh} ( 1 ) \right\}
       \approx 11.8326 $
   and considered the parameter region
   $  C'_{\omega} <  C'_{\sigma} \lesssim 30$ for $ C'_{\sigma}$.
\footnote{
   In Appendix B and C, the case of $  C'_{\sigma} =  C'_{\omega} $ and 
   the case of $  C'_{\sigma} \rightarrow 0 $ will be discussed.
                     }
In actual numerical calculations, we shall use the values
   $   C'_{\sigma} = (4/3) C'_{\omega},  (5/3) C'_{\omega}, (6/3) C'_{\omega} $,
   that is, $  C'_{\sigma} \approx 15.7768, 19.721, 23.6652 $.
It should be noticed that the vacuum mass of the hidden sector nucleon takes the same
   magnitude for different values of $C'_{\omega}$ or $C'_{\sigma}$ as discussed
   in section 2.1.
The dimensionless equation of state can be obtained numerically by
   the dimensionless energy density $\epsilon'$, Eq.(\ref{bbg}), and
   the dimensionless pressure $P'$, Eq.(\ref{bbh}).

At first we show the result of the case
   $ C'_{\sigma} = (6/3) C'_{\omega}$ in Fig.1
    in addition to that of the free fermion case.
From this figure, one can see that in the nonrelativistic region
   $ k'_{\rm F} \rightarrow 0 $ (small $\epsilon'$ region) the EOS of our model is almost the
   same as that of the free theory, and for $ k'_{\rm F} \ll 1$ the EOS of our model is softer
   than that of the free theory.
At $  k'_{\rm F} \approx 0.48$ the graph of our model intersects
   the graph of the free theory, and  for larger $ k'_{\rm F} \gtrsim 0.48$
   the EOS of our model is stiffer than that of the free theory.
At high densities $ k'_{\rm F} \gg 1 $ (large $\epsilon'$ region) the EOS of
   our model becomes $ P'/ \epsilon' \approx {\rm constant} $
   and almost the same with the EOS of the free theory.

Next we calculate the EOS in the cases of
    $ C'_{\sigma} = (4/3) C'_{\omega},  (5/3) C'_{\omega} $
    in addition to the case of $  C'_{\sigma} =(6/3) C'_{\omega} $.
All these three cases have the following same features.
In the nonrelativistic region $ k'_{\rm F} \rightarrow 0 $ (small $\epsilon'$ region) the
   three EOS's are almost the same as the EOS of the free theory, and for $ k'_{\rm F} \ll 1$
   all the three EOS's are softer than that of the free theory.
As $  k'_{\rm F} $ grows bigger, each EOS of these becomes stiffer
   than the EOS of the free theory.
At high densities
   $ k'_{\rm F} \gg 1 $ (large $\epsilon'$ region) each EOS
   of these becomes $ P'/ \epsilon' \approx {\rm constant} $
   and almost the same with the EOS of the free theory.
This fact will be explained by the analytic expression of
   $ P'/ \epsilon'$ Eq.(\ref{cbe}), i.e.,
   its numerical values of the ratio $ P'/ \epsilon'$ at high densities
   ($ k'_{\rm F} \gg 1$ or $\theta \approx \theta_f$) with
   $  C'_{\sigma} = (4/3) C'_{\omega},  (5/3) C'_{\omega} $, and
   $ (6/3) C'_{\omega} $
   are $0.3333, 0.3333,$ and $0.3333$, respectively.
Besides the above same features of the three cases,
   we will look closely at the differences between these three cases
   in the dimensionless energy density region
   $ 10^{-3} \lesssim \epsilon'  \lesssim 10^{-1} $.
In Fig.2, in order to see the dependence of EOS on the parameter 
   $  C'_{\sigma} $, we show the EOS with
   $  C'_{\sigma} = (6/3) C'_{\omega},  (5/3) C'_{\omega} $, and $ (4/3) C'_{\omega} $,
   in addition to that of the free fermion case.
%
%
\begin{figure}
     \centering
     \includegraphics[height=8.7cm]{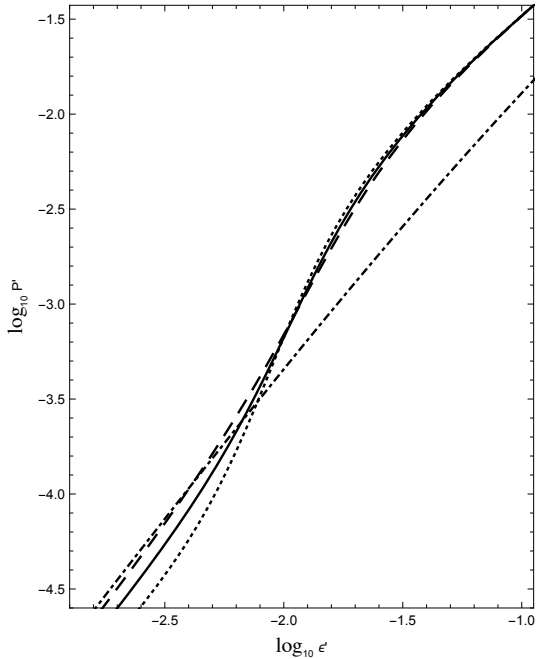}
    \caption{The graphs of EOS (dimensionless energy density $\epsilon'$ and
                  dimensionless pressure $P'$) with
                  $ C'_{\sigma} = (6/3) C'_{\omega}$ (dotted),
                  $ C'_{\sigma} = (5/3) C'_{\omega}$ (solid),
                  $ C'_{\sigma} = (4/3) C'_{\omega}$ (dashed),
                  and a free gas case (dash-dotted).
                  For all the cases except a free gas case,
                  $C'_{\omega} = (9 \pi^2 / 4 ) \left\{ \sqrt{ 2 } - {\rm arcsinh} ( 1 ) \right\}$.}
       \label{fig:2}
\end{figure}
In the nonrelativistic region $ k'_{\rm F} \ll 1$
   ($ \log_{10} \epsilon' \lesssim -2.4 $) in Fig.2, we can see that the EOS's satisfying
   $  C'_{\omega} <  C'_{\sigma} $ are softer than the EOS of the free gas model,
   and that, when \\ 
    $ \vert  C'_{\omega} -  C'_{\sigma} \vert $ becomes larger,
   the EOS becomes softer, which was already pointed out in section 3-1.
In Fig.2, one can also observe that in the range
   $ -2 \lesssim \log_{10} \epsilon' \lesssim -1 $
   the larger the value of $  C'_{\sigma} $ is, the stiffer becomes the EOS.
In order to see these differences more closely, we give Fig.3 which is the same
   with Fig.2 but magnified around $ -1.9 \lesssim \log_{10} \epsilon' \lesssim -1.3 $.
%
%
%
\begin{figure}
     \centering
     \includegraphics[height=8.7cm]{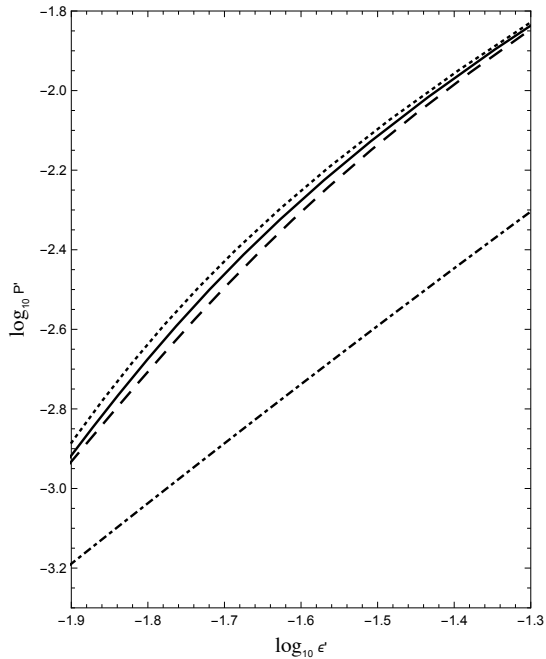}
    \caption{This graph is the same with Fig.2 but magnified around
                    $ -1.9 \lesssim \log_{10} \epsilon' \lesssim -1.3 $.}
       \label{fig:3}
\end{figure}
It seems difficult, however, to explain this characteristic by the theoretical analysis.
At the point
   $ (  \log_{10} \epsilon',   \log_{10} P' ) \approx ( -0.950, -1.427 ) $
   which corresponds to $ k'_{\rm F} =1 (\theta=1)$, EOS's with
   $  C'_{\sigma} = (6/3) C'_{\omega},  (5/3) C'_{\omega}$, and
   $ (4/3) C'_{\omega} $ coincide.
This means that the EOS does not depend on  $  C'_{\sigma} $ at the point $\theta=1$,
   which is already pointed out in the beginning of this section.
\subsection{ Numerical calculations of TOV equations }
The structure of compact stars (not rotating) with general relativity effects is
   described by the Tolman-Oppenheimer-Volkoff
   (TOV) equations \cite{rf:Tol,rf:OppVol},
\begin{equation}
   \frac{d P}{d r} = -\frac{G M \epsilon}{r^2} \left(1+ {P \over \epsilon} \right)
      \left(1+ \frac{4 \pi r^3 P}{M} \right) \left( 1- \frac{2 G M}{r} \right)^{-1},
   \label{dba}
\end{equation}
\begin{equation}
  \hskip-7.5cm  \frac{d M}{d r} = 4 \pi r^2 \epsilon,
  \label{dbb}
\end{equation}
where $P$ and $\epsilon$ denote the pressure and energy density, respectively,
   and $M(r)$ is the contained energy in a volume of radius $r$.
Giving the initial condition of $\epsilon$ at $r=0$, $\epsilon(0) \equiv \epsilon_0$
   by hand and requiring $M(0)=0$ as the initial condition, one can solve the equations 
   Eq.(\ref{dba}) and Eq.(\ref{dbb}) with the central pressure $P(0)$ obtained by EOS.
The radius of compact star $R$ is determined when the pressure $P$ becomes
   zero at the surface of the compact star, $P(R)=0$, and the total mass of the star
   is obtained by the value $M(R)$.
In the free fermion case, it has been recognized that the general character of
   the solution of the TOV equations is independent of the particle properties
   such as its mass $m_f$ and statistical weight $\gamma$
   \cite{rf:NakMor,rf:DieLabRaf}.
By taking the unit of length $a$ and the unit of mass $b$ as Eq.(\ref{aa}),
   one can transform the TOV equations to dimensionless form.
In the interacting fermion case, it can be shown that the general character of
   the solution of the TOV equations for fixed statistical weight $\gamma$
   is independent of the fermion mass $m_f$ \cite{rf:NarSchMis}.
However, in general the solution of the TOV equations with the statistical weight
   $\gamma$ can not be obtained by the use of the solution with different statistical weight
   $\gamma' (\neq \gamma)$ in the interacting fermion case.

Following Ref.\cite{rf:NarSchMis}, we shall transform the TOV equations into
   the dimensionless form.
We introduce the dimensionless quantities for the mass and the radius of the star,
\begin{equation}
   M' = \frac{M}{ \left( \frac{M^3_p}{m^2_f} \right) }, \hskip1cm
   r' = \frac{r}{ \left( \frac{M_p}{m^2_f} \right) },
  \label{dbc}
\end{equation}
where $M_p$ is the Planck mass which is expressed by the gravitational constant
   $G$, $M^{2}_p = G^{-1}$.
The TOV equations can be transformed into the dimensionless
   form \cite{rf:NarSchMis},
\begin{equation}
   \frac{d P'}{d r'} = -\frac{ M' \epsilon'}{ {r'}^2} \left(1+ {P' \over \epsilon'} \right)
      \left(1+ \frac{4 \pi {r'}^3 P'}{M'} \right) \left( 1- \frac{2 M'}{r'} \right)^{-1},
     \label{dbd}
\end{equation}
\begin{equation}
  \hskip-7.3cm    \frac{d M'}{d r'} = 4 \pi {r'}^2 \epsilon',
  \label{dbe}
\end{equation}
   where the dimensionless pressure $P'$ and dimensionless energy density $\epsilon'$
   are defined in Eq.(\ref{baj}).
The dimensionless TOV equations are solved numerically with the dimensionless
   pressure $P'$ and dimensionless energy density $\epsilon'$.
For each given initial value of $\epsilon'(0) \equiv \epsilon'_0$, the dimensionless
   radius of compact star $R'$ and dimensionless total mass of star $M'$ are obtained.
In 4.2.1 we will show the numerical result which is obtained in the case of
   $C'_{\sigma} = (6/3) C'_{\omega}$, and
   in 4.2.2 we will also give the numerical results in the cases of
   $C'_{\sigma} = (5/3) C'_{\omega}, (4/3) C'_{\omega}$ for the purpose of seeing
   the influence of $C'_{\sigma}$ on $M'-R'$ relation.
\subsubsection{ The result of $C'_{\sigma} = (6/3) C'_{\omega}$ }
In Fig.4, the numerical result of $M'-R'$ relation is represented by the dotted line in
   the case of $C'_{\omega} = (9 \pi^2 / 4 ) \left\{ \sqrt{ 2 } - {\rm arcsinh} ( 1 ) \right\} $
   and $C'_{\sigma} = (6/3) C'_{\omega}$ where dimensionless central energy density
   $\epsilon'_0$ is varied from 1.1384 to 0.000386.
%
%
%
\begin{figure}
     \centering
     \includegraphics[height=7cm]{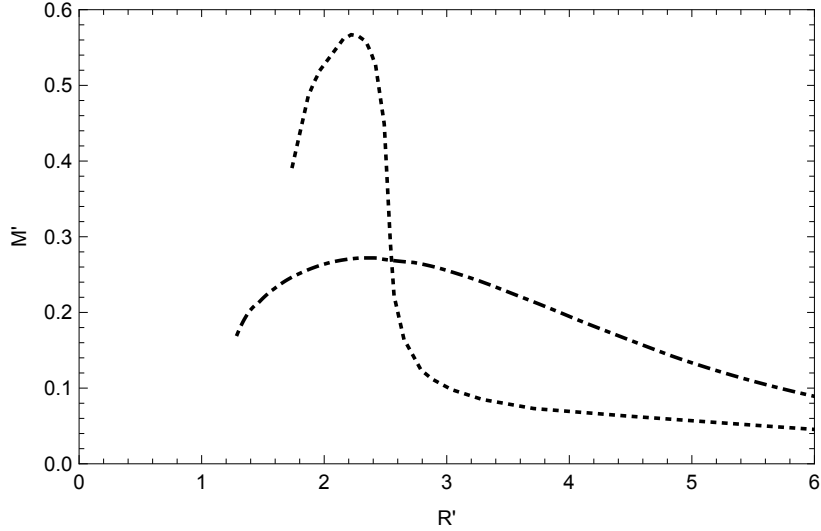}
    \caption{$M'-R'$ relations, i.e., the relations between the dimensionless star mass
                  $M'$ defined by Eq.(\ref{dbc}) and the dimensionless star radius
                  $R'$ defined by Eq.(\ref{dbc}).
                  The dotted line is the numerical result with $C'_{\sigma} = (6/3) C'_{\omega}$
                  and $C'_{\omega} = (9 \pi^2 / 4 ) \left\{ \sqrt{ 2 } - {\rm arcsinh} ( 1 ) \right\} $,
                  while the dash-dotted line is the numerical result of a free gas case.
                                                                 }
       \label{fig:4}
\end{figure}
For the star with large radius $R' \gg 1$, the smaller the central energy density
   $\epsilon'_0$ is, the larger becomes the radius $R'$.
In other words, near the center of the star the smaller the dimensionless Fermi
   momentum $k'_{\rm F}$ is, the larger becomes the radius $R'$.
For comparison, the numerical result of the free fermion case is also depicted
   by the dash-dotted line in Fig.4.
In the free fermion case, the dimensionless maximum mass of the star
   $ M'_{\rm max} = 0.272 $ is realized at the dimensionless radius
   $ R'_{\rm min} = 2.38$.
Since the EOS of the interacting fermion system is equal to that of the free fermion
   system in the limit $k'_{\rm F} \rightarrow 0$, the graph representing $M'-R'$
   relation of the interacting fermion system should coincide with that of the free
   fermion system for very large $R'$.
From the numerical calculations and the result Fig.4, we can observe the following,
\begin{description}
   \item[(a)] In the interacting fermion case, the dimensionless maximum mass
                   $ M'_{\rm max} = 0.567 $ is realized at the dimensionless radius
                   $ R'_{\rm min} = 2.24$.
                   The compactness of the star is
                   $ GM/R = M'_{\rm max} / R'_{\rm min} =0.253$.
                   The radius $ R'_{\rm min} $ of the interacting
                   fermion case is $0.94$ times as large as that of the free fermion case,
                   while the maximum mass  $ M'_{\rm max} $ of
                   the interacting fermion case is $2.1$ times as heavy as that of the
                   free fermion case. 
                   This is explained as follows.
                   When the star has its maximum mass, the dimensionless Fermi
                   momentum satisfies $k'_{\rm F} \approx O(1)$ near the center of the star,
                   where the EOS of the interacting fermion system is stiffer than that of
                   the free fermion system.
   \item[(b)] For $R' \gg 1$, the dimensionless mass $M'$ of the interacting fermion
                   system is smaller than that of the free fermion system.
                   This is explained as follows.
                   When the star has large radius $R' \gg 1$, the dimensionless Fermi
                   momentum satisfies $k'_{\rm F} \ll 1$ near the center of the star,
                   where the EOS of the interacting fermion system is softer than that of
                   the free fermion system if $  C'_{\omega} <  C'_{\sigma} $.
                   One can also observe that for $R' \gg 1$, the (absolute value of) slop
                   of the graph $M'=M'(R')$ of the interacting fermion system is smaller than
                   that of of the free fermion system.
\end{description}
\subsubsection{ Influence of $C'_{\sigma}$ on mass $M'$ and radius $R'$
                             of the star }
In Fig.4, we have represented the numerical result of $M'-R'$ relation in the
   case of
   $  C'_{\omega} = (9 \pi^2 / 4 ) \left\{ \sqrt{ 2 } - {\rm arcsinh} ( 1 ) \right\} $
   and $C'_{\sigma} = (6/3) C'_{\omega}$.
For the purpose of seeing the influence of $C'_{\sigma}$ on $M'-R'$ relation,
   the dimensionless TOV equations are also solved numerically with the same value of
   $  C'_{\omega} = (9 \pi^2 / 4 ) \left\{ \sqrt{ 2 } - {\rm arcsinh} ( 1 ) \right\} $
   but $C'_{\sigma} = (5/3) C'_{\omega}$ or $C'_{\sigma} = (4/3) C'_{\omega}$.
The results of these numerical calculations in addition to
   that of $C'_{\sigma} = (6/3) C'_{\omega}$ are shown in Fig.5.
%
%
\begin{figure}
     \centering
     \includegraphics[height=7cm]{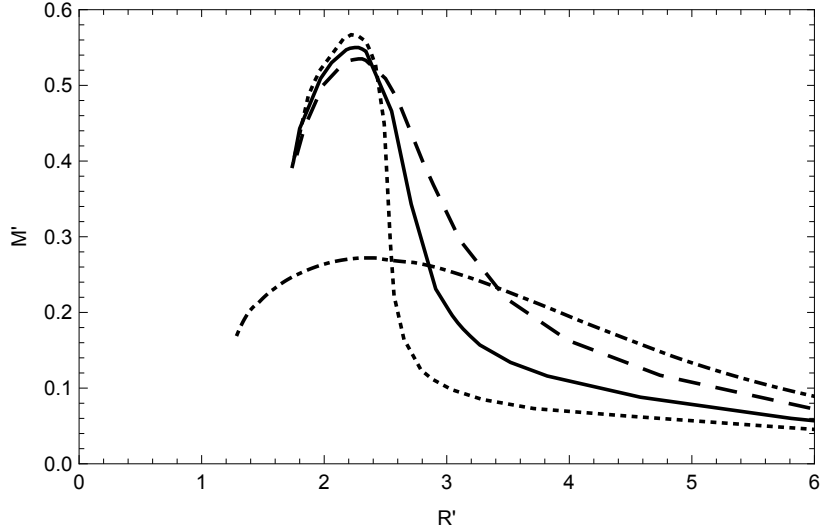}
    \caption{$M'-R'$ relations with
                  $ C'_{\sigma} = (6/3) C'_{\omega}$ (dotted),
                  $ C'_{\sigma} = (5/3) C'_{\omega}$ (solid),
                  $ C'_{\sigma} = (4/3) C'_{\omega}$ (dashed),
                  and a free gas case (dash-dotted).
                  For all the cases except a free gas case,
                  $C'_{\omega} = (9 \pi^2 / 4 ) \left\{ \sqrt{ 2 } - {\rm arcsinh} ( 1 ) \right\}$.}
       \label{fig:5}
\end{figure}
As explained previously, each graph representing $M'-R'$ relation of the
   interacting fermion system
   ($C'_{\sigma} = (6/3) C'_{\omega},  (5/3) C'_{\omega}$ and $ (4/3) C'_{\omega} $)
   should coincide with that of the free fermion system for very large $R'$.
In the case of $C'_{\sigma} = (5/3) C'_{\omega}$ ( $C'_{\sigma} = (4/3) C'_{\omega}$ ), 
   the dimensionless
   maximum mass of the star $ M'_{\rm max} = 0.550 \, (0.535) $ is realized at the
   dimensionless radius $ R'_{\rm min} = 2.25 \,(2.28)$.
From Fig.5, we can observe the following,
\begin{description}
   \item[(i)] The larger $C'_{\sigma}$ is, the heavier $ M'_{\rm max} $ becomes.
                 This is explained as follows.
                 The larger $C'_{\sigma}$ is, the stiffer EOS becomes in the range of
                 $ 10^{-2} \lesssim \epsilon' \lesssim 10^{-1} $, in which the
                 central energy density $ \epsilon'_0$ of the star lies.
   \item[(ii)] The larger $C'_{\sigma}$ is, the smaller $ R'_{\rm min} $ becomes.
   \item[(iii)] For $R' \gg 1$ the larger $C'_{\sigma}$ is, the smaller the dimensionless
                  mass $M'$ becomes.
                  This is explained as follows.
                  When the star has large radius $R' \gg 1$, the dimensionless Fermi
                  momentum satisfies $k'_{\rm F} \ll 1$ near the center of the star,
                  where the  larger $C'_{\sigma}$ is, the softer EOS becomes.
                  One can also observe that for $R' \gg 1$, the larger $C'_{\sigma}$ is,
                  the smaller the (absolute value of) slop of the graph $M'=M'(R')$ becomes.
\end{description}   
The dimensionful mass of the star $M$ and the dimensionful radius $R$ are obtained
   by dimensionless mass $M'$ and dimensionless radius $R'$
   from Eq.(\ref{dbc}) \cite{rf:NarSchMis},
\begin{eqnarray}
   M &=& 1.632 M_\odot \cdot M' \cdot \left( \frac{\rm 1 GeV}{m_f} \right)^2,
                                                                     \nonumber \\
   R &=& 2.410 \, {\rm km} \cdot R' \cdot \left( \frac{\rm 1 GeV}{m_f} \right)^2,
  \label{dbf}
\end{eqnarray}
where $m_f$ is the vacuum mass of the hidden sector nucleon.
%
%
%
%
%
%
%
\section{ Conclusion}
We studied how the EOS influences compact star made of degenerate
  hidden sector nucleons.
As the low energy effective theory of a hidden sector with a strong interaction
   like QCD, we have used a hidden sector $SU(2)$ chiral $\sigma$ model, in which
   the hidden sector vector meson $ {\omega_h}^\mu $ obtains its mass dynamically.
The mean field approximation is used and the resultant dimensionless EOS
   is determined by two dimensionless parameters  $C'_{\sigma}$ and $C'_{\omega}$.
By introducing a variable $\theta$ which depends on the dimensionless Fermi
   momentum $k'_{\rm F} = k_{\rm F}/m_f $,
   the dimensionless total energy density $\epsilon'$ and the dimensionless pressure
   $P'$ can be expressed by only one variable $\theta$, respectively.
The great advantage is that $\epsilon'$ and $P'$ are represented by
   explicit functions of this variable $\theta$, respectively, thereby
   this enables one to study characteristic features of the EOS analytically.
For example, with the parameter
   $  C'_{\omega} = (9 \pi^2 / 4 ) \left\{ \sqrt{ 2 } - {\rm arcsinh} ( 1 ) \right\} $
   which leads to $\theta=1$ when $k'_{\rm F}=1$, we can show that
   the EOS does not depend on the parameter $C'_{\sigma}$ at $\theta=1$,
   and that the EOS of our model at
   $\theta=1$ is stiffer than that of a free (hidden sector nucleon) gas model.
We consider the case of $C'_{\sigma} > C'_{\omega}$, and in actual numerical
   calculations we have fixed the value of $ C'_{\omega}$ as
   $C'_{\omega} = (9 \pi^2 / 4 ) \left\{ \sqrt{ 2 } - {\rm arcsinh} ( 1 ) \right\} \approx 11.8$
   and varied the value of $C'_{\sigma}$ such as three cases,
   $C'_{\sigma} = (6/3) C'_{\omega},  (5/3) C'_{\omega}$, and $ (4/3) C'_{\omega} $.
In the nonrelativistic region $k_{\rm F}\ll m_f \, (k'_{\rm F} \ll 1)$,
   the EOS's of our model are softer than the EOS of
   a free gas, and the larger the value of $C'_{\sigma}$ is,
   the softer becomes the EOS.
As the Fermi momentum grows bigger, the EOS's of our model become
   stiffer than the EOS of a free gas, and
   in the range $ -2 \lesssim \log_{10} \epsilon' \lesssim -1 $,
   the larger the value of $  C'_{\sigma} $ is, the stiffer becomes the EOS.
At $\theta=1$ ($k'_{\rm F}=1$ and $\log_{10} \epsilon' \approx -0.950$), EOS's with
   $C'_{\sigma} = (6/3) C'_{\omega},  (5/3) C'_{\omega}$ and $ (4/3) C'_{\omega} $
   coincide, which has been predicted analytically.
At high densities $ k'_{\rm F} \gg 1$, the EOS's of our model become
   $P' /  \epsilon'  \approx constant.$, and almost the same with the EOS
   of a free gas.

The compact star made of the hidden sector nucleons is studied by solving
   numerically the TOV equations with the EOS of our model that is determined
   by the two parameters $C'_{\sigma}$ and $C'_{\omega}$.
To begin with, we numerically calculate the case $C'_{\sigma} = (6/3) C'_{\omega}$
   and then the other cases
    $C'_{\sigma} = (5/3) C'_{\omega}, (4/3) C'_{\omega} $ are numerically studied
    in addition to $C'_{\sigma} = (6/3) C'_{\omega}$.
First, in the case of $C'_{\sigma} = (6/3) C'_{\omega}$,
   the characteristic features of the dimensionless mass $M'$ and the dimensionless
   radius $R'$ of the compact star are mentioned
   through the comparison with a free (hidden sector nucleon) gas case.
The maximum stable mass  $ M'_{\rm max}$ of compact stars
    is $2.1$ times as heavy as that of
    a free gas case, and the corresponding radius $ R'_{\rm min}$
    is $0.93$ times as large as that of a free gas case.
For somewhat large radius $R' \gg R'_{\rm min}$, the mass $M'$ of the case
   $C'_{\sigma} = (6/3) C'_{\omega}$ is lighter than that of a free gas case.
If the radius $R' $ is extremely large, the mass-radius relation ($M'-R'$ relation)
   is almost the same with that of a free gas case.
These characteristic features of $M'$ and $R'$ of the compact star are
   understandable by considering the EOS of the case
   $C'_{\sigma} = (6/3) C'_{\omega}$ as follows.
When the star has stable maximum mass, the dimensionless Fermi momentum
    $k'_{\rm F}$ satisfies $k'_{\rm F} \sim O(1)$ near the center of the star, where
    the EOS is stiffer than that of a free gas case.
When the star has somewhat large radius  $R' \gg R'_{\rm min}$, 
   $k'_{\rm F} $ satisfies $k'_{\rm F} \ll 1$ near the center of the star, where
    the EOS is softer than that of a free gas case if $C'_{\sigma} > C'_{\omega}$.
When the star has extremely large radius  $R' $, 
   $k'_{\rm F} \rightarrow 0$ near the center of the star, where
    the EOS is almost the same with that of a free gas case.
Next, the parameter $C'_{\sigma}$ is varied as
   $C'_{\sigma} = (6/3) C'_{\omega},  (5/3) C'_{\omega}$ and $ (4/3) C'_{\omega} $
   in order to see the influence of $C'_{\sigma}$ on the $M'-R'$ relation for
   the compact star.
The larger $C'_{\sigma}$ is, the heavier the maximum stable mass $ M'_{\rm max}$
   becomes.
For somewhat large radius $R' \gg R'_{\rm min}$, the larger $C'_{\sigma}$ is,
   the lighter the mass $ M'$ becomes.
These characteristic features of $M'$ and $R'$ of the compact star are
   understandable by considering the EOS whose behavior depends on the 
   parameter $C'_{\sigma}$ (we have fixed $C'_{\omega}$).
When the star has stable maximum mass $ M'_{\rm max}$, 
   the larger $C'_{\sigma}$ is,  the stiffer EOS becomes in the range of
   $ 10^{-2} \lesssim \epsilon' \lesssim 10^{-1} $ in which the dimensionless
   central energy density $ \epsilon'_0$ of the star lies.
When the star has large radius  $R' \gg R'_{\rm min}$, 
   $k'_{\rm F} $ satisfies $k'_{\rm F} \ll 1$ near the center of the star, where
    the larger $C'_{\sigma}$ is,  the softer EOS becomes.

Here, it would be desirable to discuss cooling mechanism of the star
   made of hidden sector nucleons.
In order to lower the temperature of the star, we shall consider the 
   massive hidden sector pions emission by the hidden sector nucleons.
While we dealt with completely degenerate hidden sector nucleons
   $T=0$, our result would be applicable to strongly degenerate
   hidden sector nucleons $T \neq 0$, and hence we will seek the
   necessary condition for the hidden sector nucleons $T \neq 0$ to be
   strongly degenerate.
As a concrete example, let us take the case of 
   $ C'_\sigma =(6/3) C'_\omega $ and the compact star with the
   (dimensionless) maximum mass $M'_{\rm max}$ in Fig.4, in which
   case the Fermi momentum near the center of the star is
   $k_{\rm F} \approx m_f$ and near the surface
   $k_{\rm F}\ll m_f$.
To obtain a rough estimate of the necessary condition for strongly
   degenerate we put $k_{\rm F} \approx m_f$ inside the star and
   treat the hidden sector nucleons as free particles.
It is necessary for the hidden sector nucleon to satisfy
   $ 2 \,T \ll ( \epsilon_{\rm F}-m_f ) $ so as to be strongly degenerate,
   where $ \epsilon_{\rm F}= ( k^2_{\rm F}+m^2_f )^{1/2} $ is Fermi energy.
This is because the Fermi distribution function with sufficiently low
   temperature $T \ne 0$ is different from that with a temperature
   of absolute zero in a narrow range of the energy,
   $  \epsilon_{\rm F}-T  \lesssim \epsilon \lesssim \epsilon_{\rm F}+T. $
Substituting $k_{\rm F} \approx m_f$, we obtain the necessary condition
   for strongly degenerate,
\begin{equation}
   T \ll 0.2 \, m_f.
  \label{eb}
\end{equation}
If the temperature of the star is lowered by emission of the hidden
   sector pions $\pi_h$, how its mass $m_{\pi_h}$ is constrained by
   the above necessary condition for strongly degenerate?
When a hidden sector nucleon in the state with energy $ \epsilon_{\rm F}+T $
   ($T$ is sufficiently low)
   makes a transition to a state with energy $ \epsilon_{\rm F}-T $ by
   emitting a hidden sector pion, the following condition should
   be satisfied,
   $ (\epsilon_{\rm F}-T)+m_{\pi_h} < (\epsilon_{\rm F}+T) $, or
\begin{equation}
       \frac{1}{2} \, m_{\pi_h} < T.
  \label{ec}
\end{equation}
In other words, when the temperature is lowered by emission of the
   hidden sector pions, one cannot lower the temperature $T$ below
   the value of $ m_{\pi_h}/2$.
From Eqs.(\ref{eb}) and (\ref{ec}), we obtain the constraint on the mass $m_{\pi_h}$
   of the hidden sector pion,
\begin{equation}
  m_{\pi_h} \ll 0.4 \, m_f.
  \label{ed}
\end{equation}
Although the above constraint does not put a lower limit for $m_{\pi_h}$,
   we cannot make the mass $m_{\pi_h}$ too light because it will ruin
   Big Bang Nucleosynthesis (BBN).
If, at the temperature $T_{\rm BBN} \sim 1 \, {\rm MeV}$, there exist
   relativistic particles (radiation) except those of the standard model,
   these particles will disturb BBN.
Therefore the mass $m_{\pi_h} (m_f)$ of the hidden sector pions
   (nucleons) should satisfy
\begin{equation}
 1 \, {\rm MeV} \lesssim  m_{\pi_h}, \hskip1cm  1 \, {\rm MeV} \lesssim m_f.
  \label{ef}
\end{equation}
We thus obtain the phenomenological constraint on the mass of
   the hidden sector pions,
\begin{equation}
     1 \, {\rm MeV} \lesssim  m_{\pi_h} \ll 0.4 \, m_f,
  \label{eg}
\end{equation}
and this constraint will determine the allowed region of  the term $ D \cdot \sigma_h $
   in the lagrangian (\ref{baa}) which originates from the current
   (hidden sector) quark mass.
From Eq.(\ref{eg}) the pion of the hidden sector is much lighter than
   the hidden sector nucleon.
If one introduces the Higgs mechanism to realize such an explicit chiral
   symmetry breaking case in the hidden sector, then additional
   scalars are required.
This case may potentially introduce a severe hierarchical problem.

Now we shall make two comments. 
The first comment is on the dimensionful mass $M$ of the compact star.
The dimensionful mass of the star $M$ and the dimensionful radius $R$ are
   obtained by dimensionless mass $M'$ and  dimensionless radius $R'$ \cite{rf:NarSchMis},
\begin{eqnarray}
   M &=& 1.632 M_\odot \cdot M' \cdot \left( \frac{\rm 1 GeV}{m_f} \right)^2,
                                                                     \nonumber \\
   R &=& 2.410 \, {\rm km} \cdot R' \cdot \left( \frac{\rm 1 GeV}{m_f} \right)^2,
  \label{ea}
\end{eqnarray}
where $m_f$ is the vacuum mass of the hidden sector nucleon.
Concerning Galactic searches of dark matter,
   gravitational microlensing surveys place strong upper limits on the number
   of compact objects in the Galaxy in the mass regime of
   $ (10^{-7}-30) M_\odot $ \cite{rf:Str}.
If one assumes that all dark matter is in compact stars,
   the above limits will be severe.
However, in our model the lightest hidden sector pions
   $  \mbox{\boldmath $\pi$}_h $ will also be a good candidate for cold dark matter,
   and $  \mbox{\boldmath $\pi$}_h $ will be able to exist
   outside the compact stars.
The second comment is on the choice of the value of $C'_{\omega}$.
In the numerical calculations we have fixed the parameter $C'_{\omega}$
   so as to be $y=1$ when $k'_{\rm F}$ takes a value $1$, leading to the value
   $C'_{\omega} = (9 \pi^2 / 4 ) \left\{ \sqrt{ 2 } - {\rm arcsinh} ( 1 ) \right\}$.
Although we studied analytically characteristic features of the EOS and its affects
   on the compact star to a certain extent, our technique used seems to be useful
   for this specially fixed value
   $C'_{\omega} = (9 \pi^2 / 4 ) \left\{ \sqrt{ 2 } - {\rm arcsinh} ( 1 ) \right\}$.
That is not the case.
This choice of $C'_{\omega}$ corresponds to the coupling
   $ g^2_{\sigma}/(4 \pi) \sim O(1)$.
We have used this value of $C'_{\omega}$, because the condition
   $y(k'_{\rm F}=1)=1$ may be very simple.
Another value of the parameter $C'_{\omega}$ will be chosen by requiring
    $y(k'_{\rm F}=1/2)=1$, or $f(\theta=1/2)=1$, for example, which leads to
    $C'_{\omega} \approx 110.6$ and $ g^2_{\sigma}/(4 \pi) \sim 8.8$.
Our technique used in this paper will also be applicable to the case $f(\theta=1/2)=1$
   if we again assume that the system is homogeneous and a bound state of
   hidden sector nucleon does not appear.
For example, one can show as section 4 that the EOS does not depend on
   $C'_{\sigma}$ at the point $\theta=1/2$, and that the EOS of our model at
   $\theta=1/2$ is stiffer than that of a free gas model.

It might be interesting to consider other possibility of the number of the flavor
   $N_f$ such as $N_f =3$ with statistical weight $\gamma=6$.
As discussed in section 4, in the free gas model the general character of the
   solution of the TOV equations is independent of the particle properties such as
   its mass and statistical weight $\gamma$, so that one can obtain a solution
   of the TOV equations with $\gamma=6$ by the use of a solution of the TOV
   equations with $\gamma=4$.
In the interacting model, however, it is impossible to obtain a solution of the
   the TOV equations with $\gamma=6$ by the use of a solution of the TOV
   equations with $\gamma=4$.
It will be significant to do numerical calculations of the interacting model
   with the flavor number $N_f \ge 3$.
%
%
%
%
%
%
\newpage 
\noindent{\Large\bf Appendix}
\appendix 
%
%
\section{ $y$ as a function of $ k'_{\rm F}$.}
\renewcommand{\theequation}{A.\arabic{equation}}
\setcounter{equation}{0}
With the choice of
   $ C'_{\omega} = ( 9 \pi^2 / 4 ) \left\{ \sqrt{ 2 } - {\rm arcsinh} ( 1 ) \right\} $,
   $y$ is less than $1$ when $ 0 <  k'_{\rm F} < 1$, and
   $y$ is larger than $1$ when $ 1 <  k'_{\rm F} $.
We prove this statement in this appendix.
With this value of $ C'_{\omega} $ the function $f(\theta)$ defined by Eq.(\ref{bbd})
   has the property,
\[  f( \theta )  \left\{  \begin{array}{ll}
                                 > 1    &  ( 0 < \theta < 1  )       \\
                                 =1     &  (  \theta =1 )               \\
                                  < 1   &  ( 1 < \theta < \theta_f  )      
                              \end{array}
                    \right.                      \]
For a given value of $k'_{\rm F}$, the unknown $\theta$ is obtained by solving
   the equation Eq.(\ref{bbf}), $ \theta^2 / k'^2_{\rm F} = f(\theta) $.
With the solution $\theta_{sol}$ of this equation, one can estimate the value of
   $f(\theta_{sol}) (>0)$, which satisfies
\[  f( \theta_{sol} )  \left\{  \begin{array}{ll}
                                 > 1    &  ( 0 < k'_{\rm F} < 1  )       \\
                                 =1     &  ( k'_{\rm F} =1 )               \\
                                  < 1   &  ( 1 < k'_{\rm F}  )      
                              \end{array}
                    \right.                      \]
Since $y$ is obtained by Eq.(\ref{bbc}), $y=1/ \sqrt{ f(\theta) }$,
   $y$ satisfies
\[  y   \left\{  \begin{array}{ll}
                                 < 1    &  ( 0 < k'_{\rm F} < 1  )       \\
                                 =1     &  ( k'_{\rm F} =1 )               \\
                                  > 1   &  ( 1 < k'_{\rm F}  )      
                              \end{array}
                    \right.                      \]
   for a given value of $k'_{\rm F}$.
%
%
%
%
%
%
\section{The case of $  C'_{\sigma} = C'_{\omega} $ }
\renewcommand{\theequation}{B.\arabic{equation}}
\setcounter{equation}{0}
In this appendix B, we investigate characteristic features of EOS in the case of
   $C'_{\sigma} = C'_{\omega}$ and
   $  C'_{\omega} = (9 \pi^2 / 4 ) \left\{ \sqrt{ 2 } - {\rm arcsinh} ( 1 ) \right\} $.
When $0<\theta<1$, the term
   $ \left\{ \, C'_{\omega} -  g( \theta )^2 \, C'_{\sigma}  \right\} \theta^6 $ in $P'$, Eq.(\ref{cxb}),
   becomes $  C'_{\omega} \left\{ 1 - \, g( \theta )^2 \right\} \theta^6  > 0$ because of
   $0<g(\theta)<1$, and always has positive value, indicating the EOS not to be softer.
The EOS is studied in the following three different regions.
\begin{description}
   \item[(i)]   Nonrelativistic case $ k'_{\rm F}  \ll 1  $\\
                   The $\theta$ is very small, $ \theta \ll 1 $, and from Eq.(\ref{cab}) one has 
                   $ g(\theta) \approx 1 +O(\theta^2) $.
                   We have
                   $ \left\{ \, C'_{\omega} -  g( \theta )^2 \, C'_{\sigma}  \right\} \theta^6
                      =  C'_{\omega} \left\{ \, 1 -  g( \theta )^2 \, \right\} \theta^6 \approx O(\theta^8) $.
                    From Eqs.(\ref{cxa}) and (\ref{cxb}),
                   \begin{eqnarray}
                     \epsilon'
                       &=& \frac{\gamma}{6 \pi^2 } \theta^3 + \frac{\gamma}{20 \pi^2 } \theta^5
                                + O(\theta^6),                                                 \nonumber   \\
                   P'
                     &=&   \frac{\gamma}{30 \pi^2 } \theta^5 - \frac{\gamma}{84 \pi^2 } \theta^7
                                + O(\theta^8).                                     
                       \label{hba}
                     \end{eqnarray}
                    The EOS is very similar to that of the free fermion.
   \item[(ii)]    When $ k'_{\rm F}  = 1 $.\\
                     The $\theta$ is $ \theta = 1 $, and one has $ g(\theta) = 0 $.
                     We have $ C'_{\omega} -  g(\theta)^2 \, C'_{\sigma} = C'_{\omega} $
                     and the EOS does not depend on $ C'_{\sigma}$ at $ \theta = 1 $.
                     As the discussion of the beginning of section 4,
                     the EOS of our model is stiffer compared with the free theory.
                     We have at $ k'_{\rm F}  = 1 $
                     \begin{equation}
                             \frac{ P'}{\epsilon'} \approx 0.3333.
                         \label{hbc}
                     \end{equation}
   \item[(iii)]    When $ k'_{\rm F}  \gg 1 $\\
                     The solution $\theta_f$ of the equation $f(\theta)=0$ takes the value
                     $1.1398$.
                     From Eq.(\ref{cbe}), the constant value of $P'/\epsilon'$ is $0.3333$.               
\end{description}
%
%
%
%
\section{  The case of $  C'_{\sigma} \rightarrow 0 $ }
\renewcommand{\theequation}{C.\arabic{equation}}
\setcounter{equation}{0}
If one takes a limit $C'_{\sigma} \rightarrow 0 $ formally,  it becomes $ y \rightarrow 1 $
   for any value of $ k'_{\rm F} $ from the equation of motion for $\sigma_h$.
The energy density and  pressure take the forms,
\begin{equation}
   {1 \over m^4_f} \, \epsilon
    \rightarrow    {1 \over2} \, C'_{\omega} \, n'^2_{\rm B}
          + \frac{\gamma}{ 2 \pi^2} \int^{k'_{\rm F}}_{0} d k'  k'^2 \sqrt{k'^2+1},    \\
    \label{hca}
\end{equation}
\begin{equation}
    {1 \over m^4_f} \, P
    \rightarrow    {1 \over2} \, C'_{\omega} \, n'^2_{\rm B}
          + \frac{\gamma}{ 6 \pi^2} \int^{k'_{\rm F}}_{0} d k'
             \frac{ k'^4 }{ \sqrt{k'^2+1} }.
  \label{hcb}
\end{equation}
These expressions are the same with those of Ref.\cite{rf:NarSchMis}
   if we take
\begin{equation}
   \gamma=2,
  \label{hcc}
\end{equation}
and identify
\begin{equation}
   \sqrt{  \frac{ C'_{\omega} }{2} } = y_{\rm NSM}.
  \label{hcd}
\end{equation}
%
%
%
%
%
%
%
%
%
%
%
\newpage
\end{document}